\documentclass[11pt,twoside]{article}
\usepackage{asp2010}
\resetcounters
\begin{document}

\title{UV/H$\alpha$ Turmoil}
\author{Janice C. Lee$^1$, Armando Gil de Paz$^2$, Christy Tremonti$^3$,
Robert Kennicutt$^4$ \& the Local Volume Legacy Team
\affil{$^1$Carnegie Fellow, Carnegie Observatories, 813 Santa Barbara Street, Pasadena, CA 91101; jlee@obs.carnegiescience.edu}
\affil{$^2$Departmento de Astrofisica, Universidad Complutense de Madrid, Madrid 28040, Spain}
\affil{$^3$Department of Astronomy, University of Wisconsin-Madison, Madison, WI 53706}
\affil{$^4$Institute of Astronomy, University of Cambridge, Madingley Road, Cambridge CB3 0HA, UK}
}
\begin{abstract}
A great deal of our understanding of 
star formation in the local universe 
has been built upon an extensive foundation 
of H$\alpha$ observational studies.  
However, recent work in the ultraviolet (UV) with GALEX
has shown that 
star formation rates (SFRs) inferred from 
H$\alpha$ in galactic environments 
characterized by low stellar and gas densities tend to be less
than those based on
the UV luminosity. 
The origin of the discrepancy is actively debated because one
possible explanation is that the stellar initial mass function 
is systematically deficient in high mass stars
in such environments.  In this contribution, 
we summarize our work on this topic using a dwarf galaxy
dominated sample 
of $\sim$300 late-type galaxies in the 11 Mpc Local Volume.
The sample allows us to examine the
discrepancy between H$\alpha$ and UV SFRs 
using a statistical number of galaxies
with activities less than 0.1 M$_{\odot}$ yr$^{-1}$. 
A range of potential causes for such an effect are
reviewed.  We find that
while the IMF hypothesis is not inconsistent with our observations, 
alternate explanations 
remain that must be investigated further
before a final conclusion can be drawn.
\end{abstract}

\section{The Original Motivation for this Study: Probing Starbursts in Dwarf Galaxies}

Since the late 70's and early 80's, a time when a number
of foundational papers on the measurement of star formation in external
galaxies were published (e.g., \citealt{kennicutt1983}; \citealt{gallagher1984}),
the most common way of tracing star formation 
in nearby dwarf galaxies
has been through the H$\alpha$ emission line.  
And since that time, observations of apparently
isolated dwarf galaxies undergoing starburst 
activity 
have raised the question of whether
the star formation histories of
low-mass systems are more generally characterized by bursts 
rather than modes that are quiescent and continuous 
(e.g., \citealt{gallagher1984araa}; \citealt{hunter2004}; \citealt{lee_11hugs}).

An opportunity to investigate this question in a new way
came with the launch of the GALEX satellite \citep{Martin05},
which enabled sensitive imaging of the
far ultraviolet emission (FUV; $\sim$1500 \AA) from
hundreds of dwarf galaxies in the nearby
universe \citep{Gil_etal2007,lee_10}.  
In combination with H$\alpha$ measurements
for a complete sample of galaxies,
constraints on the average duration, frequency and
amplitude of starburst episodes in the population can be computed,
since the star formation timescales probed by the two tracers
flank their expected duration (i.e.,
on the order of a dynamical time, roughly $10^8$ yrs).  
To review, H$\alpha$ nebular emission arises from 
the recombination  of gas ionized by the most massive O- 
and early-type B-stars 
($M_{\ast}\gtrsim17 M_{\odot}$). It therefore traces star formation 
over the lifetimes of these stars, which is on the order of a 
few million years.  In contrast, the UV flux primarily 
originates from the photospheres of a wider range of O- 
through later-type B-stars ($M_{\ast}\gtrsim3 M_{\odot}$), 
and thus measures star formation averaged over a longer 
$\sim10^8$ yr timescale.

Qualitatively, galaxy populations with predominantly bursty (or otherwise
discontinuous) star formation histories should imprint
a clear signature on the distribution of H$\alpha$-to-FUV
flux ratios.  
Relative to populations where the activity
is generally continuous, the average H$\alpha$-to-FUV ratio
should be depressed.  When the SFR is constant,
the ratio assumes an equilibrium value when 
the birth of stars responsible for the FUV and H$\alpha$ 
emission balance their deaths. Variations in the SFR over 
timescales on the order of $\sim$100 Myr disrupt this equilibrium. 
In the time following a burst of star formation, a deficiency 
of ionizing stars develops as they expire relative to 
lower-mass, longer-lived B-stars that produce significant amounts of 
UV emission, and the H$\alpha$-to-FUV flux ratio 
thus is lower 
(e.g., \citealt{sul00}; \citealt{sul04}; \citealt{igl04}).
For predominantly bursty populations,
determinations of the SFR based on standard linear conversions
between the luminosity and the SFR (e.g., \citealt{Kennicutt98araa}; hereafter K98) 
should also find that the
H$\alpha$-based SFRs tend be lower than those computed from the FUV,
since the conversions are derived under the assumption
of constant star formation.

Previous UV studies reported hints of such a trend for low
luminosity galaxies, and tentatively attributed 
it to a systematic 
increase in the prevalence of bursts in the recent SFHs 
of dwarf galaxies (\citealt{bellrck01}). 
However, deviations from the expected ratio 
as a function of luminosity began to appear only when 
integrated SFRs less than about 0.1 M$_{\odot}$ yr$^{-1}$ 
were probed (i.e., a few times lower than the 
SFR of the Large Magellanic Cloud), and just handfuls of 
galaxies with such low activities were included in prior 
work.  Thus, our original objectives were to
examine whether the trend persisted 
with an unbiased, statistical sample of star forming 
dwarf galaxies, and if so, to use the average H$\alpha$-to-FUV 
ratios to produce new constraints on the temporal
variability of the integrated activity.

Analysis with our dataset indeed revealed that the dwarf galaxies 
had systematically lower H$\alpha$-to-FUV 
values than more luminous, higher surface brightness
spiral galaxies.  This seemed to confirm
earlier suggestions of generally bursty star formation 
in low mass systems, and standard population
synthesis modeling showed that bursts which 
elevate the SFR of galaxies by a factor of $\gtrsim$100 
for a 100 Myr duration would reproduce
the observed ratios.  However, such large amplitudes
appeared to be in conflict with other independent
determinations of the recent star formation history (SFH), which
showed that more modest enhancements (factor of
$\sim$5) were typical.  We therefore were led to consider
a wider range of potential causes for the low observed
H$\alpha$-to-FUV ratios in dwarf galaxies.
Here, we distill the main points of our analysis, 
as reported in \citet{LeeGildePaz_etal2009}.
Discussion of this topic also appears in a number
of other contributions in these Proceedings, by
Meurer; Boselli et al.; Eldridge; Johnson; Calzetti;
and Pflamm-Altenburg; among others.

\section{Data}
Our study is based upon data collected by the GALEX 11HUGS
(11 Mpc H$\alpha$ UV Galaxy Survey)
and Spitzer LVL (Local Volume Legacy) programs. 
The sample is dominated by dwarf galaxies, 
and is thus ideal for studying the nature of 
systems with low SFRs. Integrated H$\alpha$, 
UV, and mid- to far-IR flux catalogs are 
published in \citet{K08_ha}, and \citet{lee_10}, 
\citet{dale+09}, respectively. Details on the 
sample selection, observations, photometry, and 
general properties of the sample are provided in 
those papers. A brief summary of the dataset 
is given here.

Our total Local Volume sample contains 436 objects.  
Galaxies are compiled from existing catalogs 
(as described in \citealt{K08_ha}), and the selection 
is divided into two components. The primary component
of the sample aims to be as complete as possible 
in its inclusion of known nearby star-forming galaxies 
within given limits.  It consists of spirals and irregulars 
brighter than B = 15 mag within 11 Mpc that avoid the 
Galactic plane ($|b|>$20$^{o}$).  These bounds 
represent the ranges within which the original 
surveys that provided the bulk of our knowledge 
on the Local Volume galaxy population have been shown 
to be relatively complete,
while still spanning a large enough volume to probe 
a representative cross section of star formation properties. 
The secondary component of the sample consists of galaxies 
that are within 11 Mpc and have available H$\alpha$ 
flux measurements, but fall outside one of the 
limits on brightness, Galactic latitude, or 
morphological type.  It is a composite of 
targets that were either observed by our group as 
telescope time allowed, 
or had H$\alpha$ fluxes published in the literature. 
Subsequent statistical tests, as functions of 
B-band apparent magnitudes and HI fluxes (compiled from
the literature), show that 
the subset of galaxies with $|b|>20^{\circ}$ 
is relatively
complete to $M_B\lesssim-15$ and $M_{HI}> 2 \times 10^8$ 
M$_{\odot}$ at the edge of the 11 Mpc volume \citep{lee_11hugs}.

Through a combination of new narrowband H$\alpha$+[NII] 
and $R$-band imaging, and data compiled from the literature, 
integrated H$\alpha$ fluxes are available for over 90\% of the total sample. 
The new narrowband imaging 
obtained by our group was taken at 1-2 m class telescopes 
in both hemispheres, and reached relatively deep point source 
flux and surface brightness limits of $\sim$2 $\times$ 10$^{-16}$ 
ergs cm$^{-2}$ s$^{-1}$ and $\sim$4 $\times$ 10$^{-18}$ ergs cm$^{-2}$ 
s$^{-1}$ arcsec$^{-2}$, respectively.

Subsequent GALEX UV imaging primarily targeted 
the $|b|>30^{\circ}$, $B<15.5$ subset of the sample.  
The more restrictive latitude limit was imposed to 
avoid excessive Galactic extinction and fields 
with bright foreground stars and/or high background 
levels for which observations would be prohibited due 
to GALEX's brightness safety restrictions.  Deep, 
single orbit ($\sim$1500 sec) imaging was obtained 
for each galaxy, following the strategy of the 
GALEX Nearby Galaxy Survey \citep{Gil_etal2007}.  GALEX 
observations for a significant fraction of the remaining 
galaxies beyond these limits were also taken by other GI
programs.
Overall, GALEX data are available for $\sim90$\% of the sample.  

Finally, Spitzer IRAC mid-infrared and MIPS far-infrared 
imaging was also obtained for the $|b|>30^{\circ}$, $B<15.5$ 
subset of the sample through the Local Volume Legacy program. 
The far-infrared photometry provide attenuation corrections for our analysis.

\section{Results}

\begin{figure}[t!]
\includegraphics[scale=0.35,bb=-200 144 592 718]{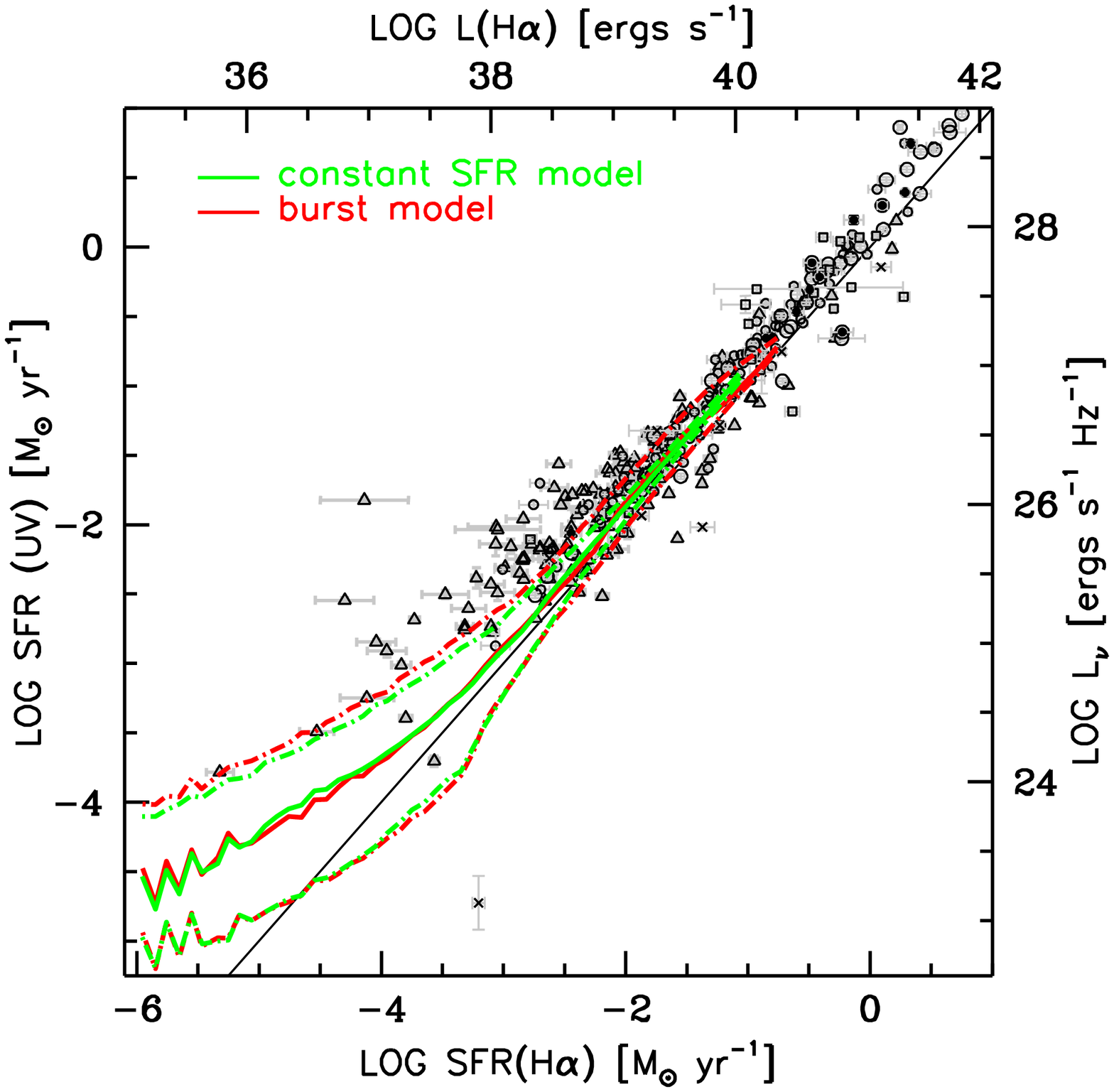}

\includegraphics[scale=0.35,bb=-200 144 592 718]{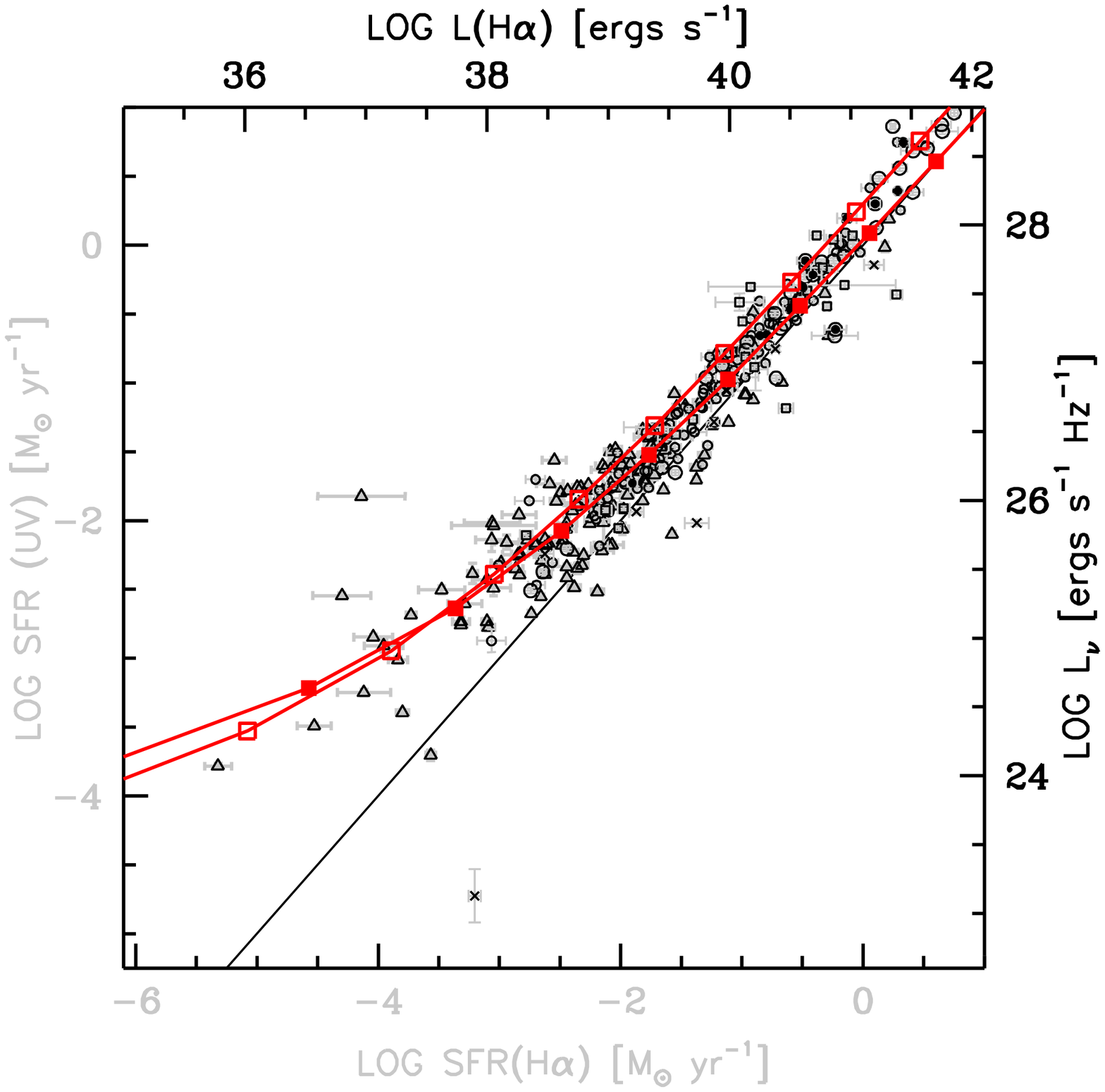}
\caption{Comparison of FUV and H$\alpha$ luminosities (corrected for internal dust attenuation),
with axes indicating the corresponding H$\alpha$ and FUV SFRs,
computed from the linear conversion recipes of K98.
The solid line represents a one-to-one correspondence between the SFRs.
The top panel show the data with predictions from models
which perform population synthesis by random sampling of
the IMF \citep{tremonti07}.  The median-predicted values 
(solid line) are shown along with
values at the 2.5 and 97.5 percentile points (dotted lines).
Predictions from IGIMF model of 
\citet{kroupa+weidner2003}, as computed by \citet{PWK07} 
are shown with the data in the bottom panel.
The bottom and left hand axes are shown in gray 
to signify that the K98 SFR scales would not be
valid at low SFRs under the assumptions of the IGIMF model.
\label{fig1}}
\end{figure}

\begin{figure}[t!]
\includegraphics[scale=0.55,bb=0 400 592 718]{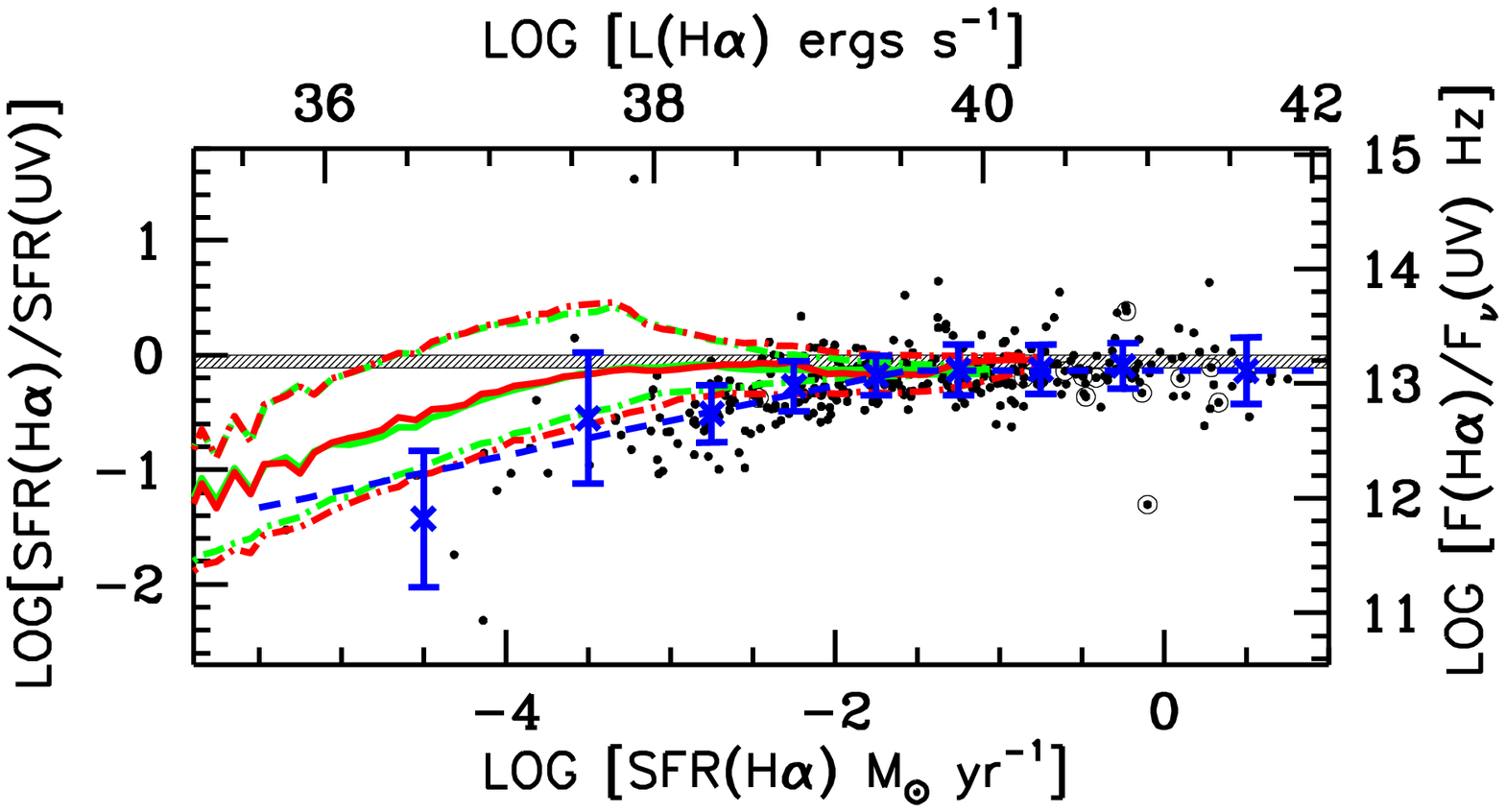}
\includegraphics[scale=0.55,bb=0 400 592 718]{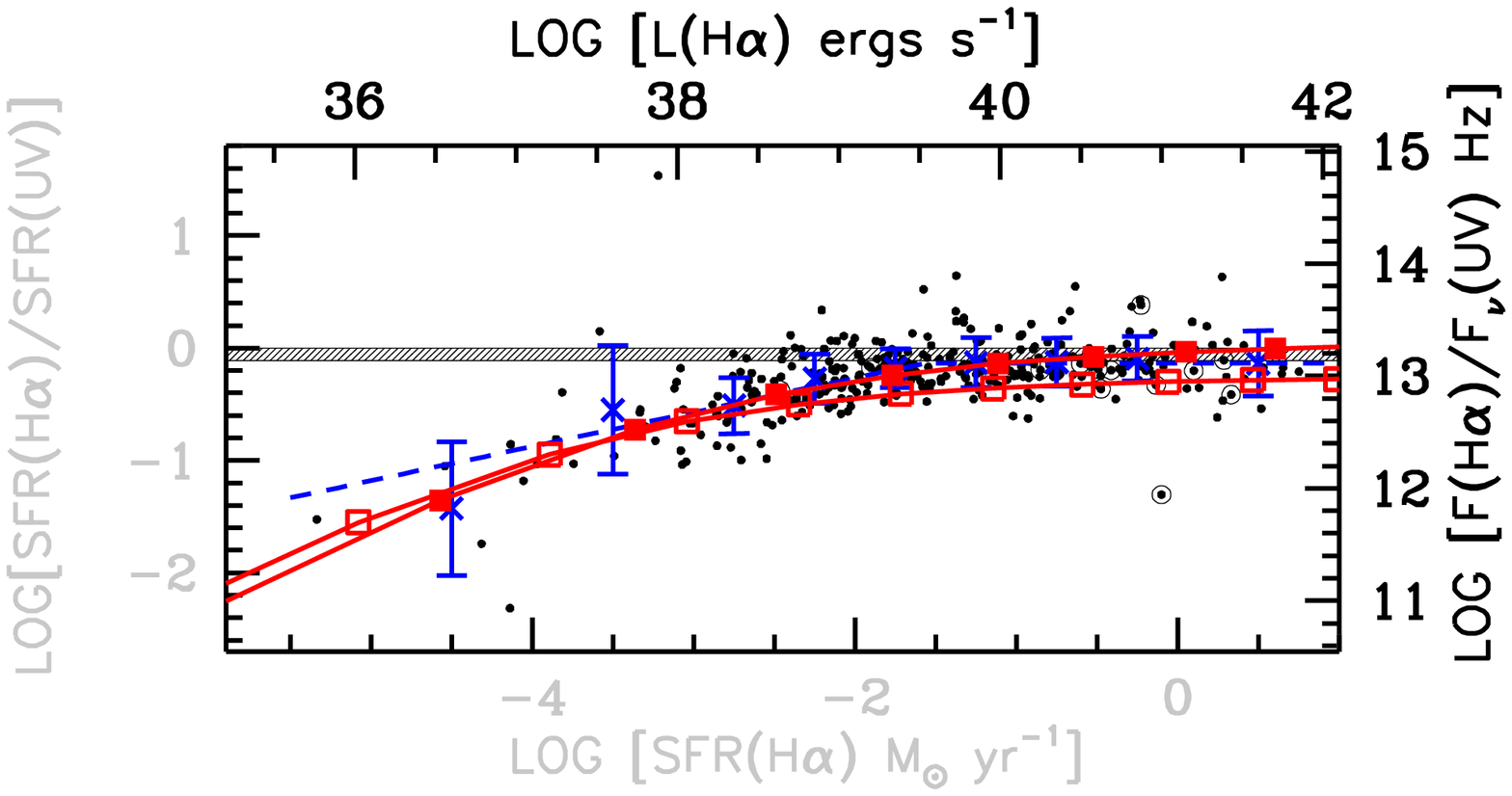}
\caption{Analogous to Figure 1, but for the ratio of H$\alpha$-to-FUV SFRs/luminosities 
(corrected for internal dust attenuation)
as a function of the H$\alpha$ SFR/luminosity.
The shaded band represents the range of H$\alpha$-to-FUV 
ratios predicted by commonly used stellar population models 
for constant star formation.  The dashed line shows a linear least squares fit 
to the data.  Circled points represent galaxies where the H$\alpha$ 
flux may be underestimated because the narrowband imaging did not wholly enclose the galaxy.
The top panel shows the deviations expected from random sampling
of the IMF, while the bottom panels shows the predictions
of the IGIMF model.
\label{fig2}}
\end{figure}

Our main results are shown in Figures \ref{fig1} and \ref{fig2}.
Figure \ref{fig1} plots the FUV luminosity against the H$\alpha$
luminosity, while Figure \ref{fig2} shows the H$\alpha$-to-FUV
flux ratio as a function of the H$\alpha$ luminosity.  
A systematic decrease in the H$\alpha$-to-FUV
ratio with decreasing luminosity is evident.  
At high SFRs
($\gtrsim$ 0.1 M$_{\odot}$ yr$^{-1}$), the ratio 
is constant, and the value of SFR(H$\alpha$)/SFR(FUV) is $\sim$0.75.
Within the uncertainties, this value is consistent with
a zero offset from the expected
equilibrium value for constant star formation. 
However, by SFRs of about 0.003 
M$_{\odot}$ yr$^{-1}$, the average H$\alpha$-to-FUV flux ratio 
is lower by a factor of two, and 
at the lowest SFRs probed (10$^{-4}$ M$_{\odot}$ yr$^{-1}$), 
the average deviation
is about a factor of ten.  

Best-effort attenuation corrections have been applied to 
the values plotted in Figures \ref{fig1} and \ref{fig2}
(i.e., based on Balmer decrements and total infrared to FUV ratios
when available, and on empirical scaling relationships otherwise).  
However, the trend is already evident prior to these corrections; it thus
should be robust to the effects of dust since the 
FUV should be more severely attenuated than the nebular H$\alpha$ emission.

\section{Understanding the Systematic Decline in L(H$\alpha$)/L(FUV)}

Again, our analysis confirms previous indications of
systematically lower H$\alpha$-to-FUV ratios in dwarf
galaxies based on small datasets ($\sim$ 20 galaxies
with SFRs $\lesssim$ 0.1 M$_{\odot}$ yr$^{-1}$), and improves
the sampling of this regime by an order of magnitude.  
Our result is also corroborated by a number of concurrent, 
independent studies (though with fewer numbers of dwarfs),
and the trend has also been reported as a function of 
decreasing optical surface brightness and stellar 
mass \citep{Meurer_etal2009, Hunt2010, bos09}.

As discussed above, our initial assumption was that
such a trend would be an indication of an increased
frequency of starburst activity in the dwarf galaxy
population, and that modeling the decline of the
average ratio would yield new constraints on the
characteristic durations, frequencies and amplitudes
of the bursts.  Examination of a Starburst99 model
grid spanning a range of burst parameters
computed by \citet{igl04} showed that the
factor of two offset observed at SFR$\sim$0.003 M$_{\odot}$ yr$^{-1}$
could be reproduced by bursts with amplitudes
of $\gtrsim$100 lasting for 100 Myr.  However, 
such large amplitude bursts appear to be in conflict
with other observational constraints.  The most direct 
constraints are provided by studies which reconstruct 
star formation histories from resolved observations of 
stellar populations in nearby low-mass systems
(e.g., \citealt{wei08}; \citealt{mcq09}; \citealt{mcquinn10}).
The typical burst amplitudes found in such studies
range from a few to $\sim$10, 
an order of magnitude smaller than the factor of 100 bursts 
required by the models to depress the H$\alpha$-to-FUV 
flux ratios by a factor of two.  
The 11HUGS sample itself also provides a statistical constraint 
on the average dwarf galaxy starburst amplitude via the 
ratio of fraction of star formation (as traced by H$\alpha$) 
concentrated in starbursts to their number fraction \citep{lee_11hugs}.  
The high degree of statistical completeness of 11HUGS makes this 
calculation possible for galaxies with $M_B<-15$, and again, 
the burst amplitude is found to be relatively modest ($\sim$4).

We were thus led to consider a wider range of explanations
for the trend, the most viable of which are:
\begin{quote}
(1) a leakage of ionizing photons from the low
density environments which are characteristic of the galaxies
where low H$\alpha$-to-FUV ratios are observed;\\
(2) low probabilities of forming ionizing
stars when the integrated SFR of a galaxy is low;\\
(3) an IMF which is systematically deficient
in the highest mass stars in low density environments.
\end{quote}

The first possibility is notoriously difficult to
observationally constrain, though the few existing 
studies of Lyman continuum escape from 
present-day galaxies have all found upper limits 
$\lesssim$10\% (e.g., \citealt{claus95}; \citealt{nils06}).  These detection experiments have 
been performed on starbursting galaxies where 
leakage is thought to most likely occur.  Although
it is difficult to conclusively rule out this
possibility, given the reported upper-limits, it
seems improbable that ionizing photon losses 
could approach the levels required to explain
the factor of two discrepancies observed, particularly
in the quiescent, gas rich dwarfs that dominate our sample.

A related explanation is that low gas densities
lead to extremely diffuse
H$\alpha$ emission that falls beneath
the detection threshold of standard narrowband observations \citep{Hunt2010}.
In this case, the ionizing photons 
travel far from their parent HII regions,
but do not to escape the galaxy 
entirely.
To check whether this could be an issue
for our data, we performed the following simple calculation.
The potentially missing H$\alpha$ luminosity is computed assuming 
that the H$\alpha$ star formation rate (SFR) 
should be consistent with the SFR inferred 
from the UV continuum flux.  The surface brightness 
of this emission is then estimated with the additional 
assumption that the nebular flux is uniformly 
distributed over the area of the UV disk.  
We found that, under these assumptions,
the missing H$\alpha$ should have been unambiguously 
detected in our observations.  
However, if the nebular emission instead extends to 
twice the area of the UV disk 
(this generally encloses the periphery
of the HI gas distribution, and thus serves
as a limiting case), then the emission will
fall below our sensitivity limits.
This issue is one that requires further examination.
To do this, we recently obtained
deep narrowband ($\sim$8\AA) imaging of a few
dwarf galaxies with the
Maryland-Magellan Tunable Filter on the 
Inamori Magellan Areal Camera and Spectrograph (IMACS) 
at the Magellan 6.5m.  The data will allow us
to observe H$\alpha$ down to $\sim$4
$\times$10$^{-18}$ ergs s$^{-1}$ cm$^{-1}$ arcsec$^{-2}$ ---
sensitive enough to detect the surface brightnesses
expected if the ``missing H$\alpha$" is in fact spread over 
the entire area of the gas disk.
 
The second possibility is that random sampling of a universal IMF 
in the regime of ultra-low integrated SFRs, may lead to an apparent
deficiency or absence of H$\alpha$, even when there is on-going
star formation, because the probabilities of
forming high-mass ionizing stars are low.  We use Monte Carlo
simulations to model the impact
of such stochasticity on the H$\alpha$-to-FUV flux ratio,
and compare the results to the data in the top panels of
Figures \ref{fig1} and \ref{fig2}.
The median-predicted values (solid line) are shown along with
values at the 2.5 and 97.5 percentile points (dotted lines). 
While stochasticity clearly does have an effect,
its impact does not appear to be large enough to explain
the observed trend by itself.

Finally, we also considered the possibility that 
the IMF is systematically deficient in the highest mass
stars in the low density environments of dwarf galaxies.
In particular, we compare the model of the Integrated
Galactic IMF (IGIMF; \citealt{kroupa+weidner2003}; also see
Pflamm-Altenburg; and Weidner, in these Proceedings) to
the data in the bottom panels of Figures \ref{fig1} and \ref{fig2}.
The model assumes that the most massive star that
can form depends on its parent cluster's mass 
in a deterministic (rather than a probabilistic) manner,
while the mass of the most massive cluster that
can form is dependent on a galaxy's integrated star formation
rate.  This model is able to reproduce the trend
observed in our data.

\section{Discussion and Next Steps}

While the consistency between the IGIMF model
and the data is intriguing, alternate 
explanations have not yet been
explored to the point where they
can be conclusively eliminated.
Clearly, the true fate of the ionizing photons
in low density gas will be difficult to 
determine, though we hope that our new
ultra-deep tunable filter imaging will
provide new insight on this issue.
There was general consensus at the meeting
that this must be investigated further. 

Whether the starburst scenario, or
some class of fluctuating star formation histories,
can be ruled out as the cause of the
low H$\alpha$-to-FUV flux ratio was a question that
spawned much greater debate (e.g., contributions
by Boselli et al. and Meurer in these Proceedings).  
Our position has
been that the modeling of star formation histories
warrants further examination, particularly in 
conjunction with the modeling of stochastic sampling
of the IMF.  Although we currently find that
the large burst amplitudes required to reproduce
the trend are in conflict with other
observational constraints, stochasticity may 
amplify the effects of bursty or non-uniform "flickering" 
SFHs on the H$\alpha$-to-FUV ratio, and reduce
the required amplitudes.  Fortunately, a new
generation of stellar population
synthesis models that incorporate random sampling
of a standard IMF 
has been recently developed
by several groups (e.g., see contributions
by Eldridge; Fumagalli; and Cervi\~{n}o in these Proceedings).
Such models are much better suited for 
probing the properties of systems
where the IMF may not be fully sampled,
compared to existing models such as Starburst99
which use fully-populated
$\sim$10$^5$ M$_{\odot}$ quanta as the basis 
for the synthesis.  
Overall, it seems possible that some combination 
of all of the effects described above 
may conspire to produce the observed systematic.

Although the origin of the trend is currently unclear,
there is one reasonable conclusion that can
be drawn from our study: that
FUV luminosities provides more accurate SFRs for 
individual galaxies with low total SFRs 
and low dust attenuations.  The
FUV emission should be less prone to stochastic effects 
from sparse sampling of the upper end of the 
IMF, and to possible uncertainties in the fate of 
ionizing photons.  With the demise of the GALEX
FUV detector however, there is no longer a 
facility capable of obtaining such data,
and for the foreseeable future, 
new measurements of the SFR in dwarf galaxies
will need to be based on H$\alpha$ emission.
The role of the new stochastic stellar population synthesis
models is thus critical, as they will be necessary
to interpret the most likely value and possible range of
the SFR that a given H$\alpha$ luminosity corresponds
to.

\acknowledgements 
JCL gratefully acknowledges support from the Hubble and Carnegie 
Postdoctoral Fellowship Programs.  She is also thankful
for funding from the GALEX project, 
which enabled not only the science presented
here, but also the organization of the overall Up2010
workshop.

\bibliography{up2010refs}
\end{document}